\documentclass[]{aa} 

\usepackage{graphicx}
\usepackage{natbib}
\bibpunct{(}{)}{;}{a}{}{,} 
\usepackage{txfonts}

\begin{document}

\title{Catching the fish -- Constraining stellar parameters for TX\,Psc using spectro-interferometric observations\thanks{Based on observations made with ESO telescopes at  Paranal Observatory under program IDs 74.D-0601, 60.A-9224, 77.C-0440, 60.A-9006, 78.D-0112, 84.D-0805}}

\author{D. Klotz\inst{1} \and C. Paladini\inst{1} \and J. Hron\inst{1} \and B. Aringer\inst{1} \and S. Sacuto\inst{1, 2} \and  P. Marigo\inst{3} \and T. Verhoelst\inst{4,5}}

\institute{Department of Astrophysics, University of Vienna, T\"urkenschanzstrasse 17, A-1180 Vienna\\
              \email{daniela.klotz@univie.ac.at}
    \and
     Department of Physics and Astronomy, Division of Astronomy and Space Physics, Uppsala University, Box 516, 75120, Sweden
     \and
     Department of Physics and Astronomy {\em G.\ Galilei}, University of
Padova, Vicolo dell'Osservatorio 3, I-35122 Padova, Italy
     \and
     Belgian Institute for Space Aeronomy (BIRA-IASB), Ringlaan-3-Avenue Circulaire, B-1180 Brussels, Belgium
     \and
     Instituut voor Sterrenkunde, KULeuven, Celestijnenlaan 200D, 3001 Heverlee, Belgium
               }

\date{Received ; accepted}

\abstract
   {Stellar parameter determination is a challenging task when dealing with galactic giant stars. The 
   combination of different investigation techniques has proven to be a promising approach. 
   We analyse archive spectra obtained with the  Short-Wavelength-Spectrometer (SWS) onboard 
   of ISO, and new interferometric observations from the Very Large Telescope MID-infrared 
   Interferometric instrument (VLTI/MIDI) of a very well studied carbon-rich giant: TX\,Psc.}
   {The aim of this work is to determine stellar parameters 
   using spectroscopy and interferometry. The observations are used to constrain the model 
   atmosphere, and eventually the stellar evolutionary model in the region where the tracks map 
   the beginning of the carbon star sequence. }	
   {Two different approaches are used to determine stellar parameters: 
   (i) the `classic' interferometric approach where the effective temperature is fixed by using the 
       angular diameter in the $N$-band (from interferometry) and the apparent bolometric magnitude; 
   (ii) parameters are obtained by fitting a grid of state-of-the-art hydrostatic models to spectroscopic and 
        interferometric observations.}
   {We find a good agreement between the parameters of the two methods. 
   The effective temperature and luminosity clearly place TX\,Psc in the 
   carbon-rich AGB star domain in the H-R-diagram. Current evolutionary tracks suggest that TX\,Psc became a 
   C-star just recently, which means that the star is still in a `quiet' phase compared to the subsequent strong-wind 
   regime. This is in agreement with the C/O ratio being only slightly larger than 1.
   }
   {}
   

\keywords{Stars: AGB and post-AGB - Stars: atmospheres - Stars: carbon - 
                 Stars: fundamental parameters - Techniques: interferometric - 
                 Techniques: spectroscopic}
\authorrunning{Klotz et al.}          
\titlerunning{Constraining stellar parameters for TX\,Psc}

\maketitle


\section{Introduction}
  The asymptotic giant branch (AGB) is the late evolutionary stage of low- to intermediate- mass 
  stars ($1-8$\,$M_{\sun}$). On the early-AGB the carbon-to-oxygen-ratio is smaller than one. After 
  several thermal pulses, the atmospheres of objects with masses between $1-4$\,$M_{\sun}$ will 
  very likely turn from oxygen-rich into carbon-rich because of the third dredge-up \citep{iben83}. 
  Good estimates of stellar parameters are needed for a profound understanding of the evolution of 
  this stage. Their determination is a challenging task because of the complexity of the atmospheres 
  of these objects. It is demonstrated that the combined use of spectroscopic and 
  interferometric observing techniques can efficiently help to ascertain stellar parameters 
  \citep[e.g.][]{wittkowski01,wittkowski08,wittkowski11,neilson08,paladini11,sacuto11,martividal11}. 
  At the same time these observations provide constraints for existing model atmospheres: 
  e.g.\,COMARCS \citep{aringer09},  \citet{hoefner03}, PHOENIX \citep{hauschildt99}, ATLAS \citep[e.g.][]{kurucz93}, 
  CODEX \citep{ireland08,ireland11}.\\
  TX\,Psc is one of the brightest and closest carbon-rich AGB stars. It is listed as an irregular variable in 
  the General Catalogue of Variable stars \citep{samus09} with a mean brightness of 
  $\sim$5\,mag and a peak-to-peak amplitude of 0.4\,mag in the $V$-band \citep{jorissen11}. Distance 
  estimates range from 275 to 315 pc \citep{vanleeuwen07,claussen87,bergeat05}. Different ISO/SWS 
  spectra of TX\,Psc \citep{jorgensen00,gautschy04} show that there is a difference in the 3\,$\mu$m 
  feature between 1996 and 1997. With the help of plane-parallel, hydrostatic models 
  \citet{jorgensen00} interpret this difference as a change in temperature of $\sim$100\,K. The 
  photometry and spectra were successfully modeled by \citet{gautschy04} with dust-free dynamical 
  models that reproduce the region between $1-5$\,$\mu$m. They claim that the region between 
  $8-9$\,$\mu$m is affected almost solely by CS. 
  The authors suggest that observations from $8-9$\,$\mu$m show the deep 
  photosphere, while the expected absorption of HCN and 
  C$_2$H$_2$ originating from the higher layers is not observed. As this object is almost (carbon) 
  dust-free we may infer it became a carbon star quite recently. Therefore, this star is a perfect candidate 
  to constrain the region where the transition from oxygen- to carbon-rich occurs.  \\
  In this work we present a study of the atmosphere and a stellar parameter determination for TX~Psc. We 
  combine spectro-interferometric observations of VLTI/MIDI and spectroscopic observations from ISO/SWS 
  and compare them to geometric, hydrostatic, and evolutionary models.\\ 
  A description of observations and data reduction is given in 
  Sect.\,\ref{obs-reduction}. Models and approaches to derive synthetic observables are 
  presented in Sect.\,\ref{model-description}. The stellar parameters are derived in Sect.\,\ref{parameterdetermination} 
  and compared to state-of-the-art evolutionary tracks in Sect.\,\ref{evoltracks}. A summary of the results is 
  given in Sect.\,\ref{conclusion}.


\section{Observations and Data Reduction}
\label{obs-reduction}
  Sections\,\ref{observations} and \ref{observations-iso} discuss the 
  interferometric and spectroscopic observations obtained with VLTI/MIDI and 
  ISO/SWS, respectively. Section\,\ref{variability} discusses possible cycle-to-cycle and intra-cycle variability 
  of the data.

  \subsection{MIDI visibilities and spectra}
  \label{observations}
	\begin{figure}
		\centering
		\includegraphics*[width=7.cm,bb=90 375 463 703]{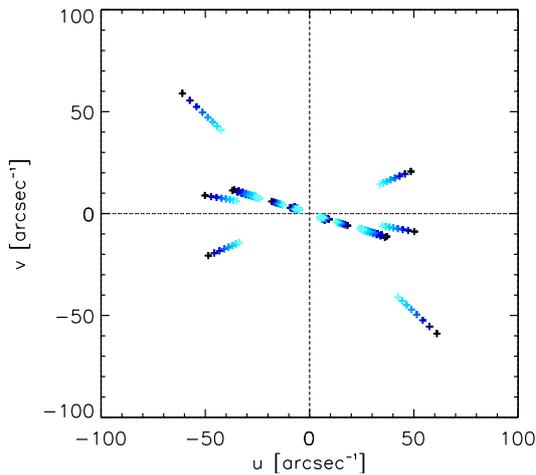}
		\caption{\label{uv-coverage}$N$-band spectrally dispersed $uv$-coverage of the MIDI observations 
		of TX\,Psc. Colour levels range from 
		8-11.5\,$\mu$m (black to blue, respectively) with a step size of 0.5\,$\mu$m. North is up and East 
		is left.}
	\end{figure}
    \object{TX Psc} was observed in 2004 with the 8.2\,m Unit Telescopes and in 2005, 2006 and 2011 with 
    the 1.8\,m Auxiliary Telescopes of the Very Large Telescope Interferometer MIDI \citep{leinert03}. MIDI 
    covers the $N$-band and provides spectrally dispersed visibilities, differential phases and fluxes 
    (resolution $\mathrm{R}=230$ for observations in 2004, $\mathrm{R}=30$ for observations in 2005, 
    2006 and 2011). \\
    The journal of available MIDI observations is given in Table~\ref{journal} (electronic version only). 
    The $uv$-coverage is plotted in Fig.\,\ref{uv-coverage}.  \\
      \onltab{1}{
    \begin{table*}[htdp]
	    \caption{\label{journal}Journal of MIDI observations of TX\,Psc sorted for projected baseline length. }
	    \centering
	    \vspace{0.5cm}
	    \begin{tabular}{llcccccccccc}
	    \hline
	    \hline
	    \# & Object & Date/Time & Configuration & Proj.Base & Proj.Angle & Obs. seeing & Airmass \\
	     & & & & [m] & [$^\circ$]&[\arcsec]\\
	    \hline
	    1	&TXPSC			 	 	& 2006-08-28 02:50 			& E0-G0 		& 9.4 	& 81 		& 1.41	& 1.718\\    
		\hline       	
		2	&\textbf{TXPSC}\tablefootmark{a} 	& 2006-10-19 00:19		  	& E0-G0  		& 11.9 	& 70 		&  		& 1.387\\ 
			&HD48915 				& 2006-10-19 08:26			& \ldots 		& \ldots 	& \ldots 	& 0.57	& 1.325\\ 
		\hline
		3	&\textbf{TXPSC} 					& 2006-10-18 05:46			& E0-G0  		& 13.2 	& 66		& 0.89 	& 1.659\\ 
			&HD48915 				& 2006-10-18 06:57			&  \ldots 		& \ldots 	& \ldots 	& 0.87	& 1.278\\ 
		\hline
		4	&TXPSC			 	 	& 2005-06-29 08:55 			& E0-G0 		& 13.7 	& 82 		& 0.54 	& 1.238 \\
		\hline 
		5	&TXPSC			 		& 2010-09-07 04:13			& E0-G0		&14.4	& 73		& 0.93	& 1.194\\
		\hline 
		6	& TXPSC			 		& 2006-08-28 05:48	 		& E0-G0		& 15.6 	& 73 		& 1.30 	& 1.136\\
		\hline
		7 	& TXPSC			 		& 2009-11-15 02:18 			& E0-G0 		& 15.7 	& 72 		& 1.26 	& 1.225 \\
		\hline  
		8	&\textbf{TXPSC}\tablefootmark{a}  	& 2006-10-18 03:18			& E0-G0  		& 16.0 	& 73 		& 1.00 	& 1.149\\
			&HD48915 				& 2006-10-18 06:57			&  \ldots 		& \ldots 	& \ldots 	& 0.87	& 1.278\\ 
		\hline
		9	&TXPSC			 	  	& 2006-09-20 02:28			& D0-G0  		& 25.1 	& 71 		& 1.21 	& 1.327\\
		\hline
		10	&TXPSC			 	  	& 2006-09-20 03:23			& D0-G0 		& 29.0 	& 73 		& 1.90  	& 1.189 \\ 
		\hline    
		11	&TXPSC			 	 	& 2006-10-16 02:12			& H0-G0 		& 30.5 	& 73 		& 0.67 	& 1.150\\
		\hline
		12	& TXPSC			 	 	& 2009-11-16 02:13 			& H0-G0 		& 31.0 	& 72 		& 0.95 	& 1.219 \\
		\hline
		13	&\textbf{TXPSC}					& 2006-09-20 05:47			& D0-G0 		& 31.6 	& 72 		& 1.64 	& 1.201\\
			&HD20720 				& 2006-09-20 09:48			& \ldots 		& \ldots 	& \ldots 	& 0.89	& 1.095\\ 
		\hline
		14 	& TXPSC			 	 	& 2009-11-16 01:20 			& H0-G0 		& 32.0 	& 73 		& 1.36 	& 1.146 \\
		\hline
		15	&TXPSC			 	 	& 2006-09-21 01:48			& K0-G0  		& 44.0 	& 68 		& 1.01	& 1.489 \\  
		\hline
		16	&\textbf{TXPSC}\tablefootmark{a} 	& 2006-09-21 03:32			& K0-G0   		& 59.3 	& 73 		& 1.02 	& 1.169\\
			&HD48915				& 2006-09-21 08:36			& \ldots 		& \ldots 	& \ldots 	& 1.51	& 1.309\\    
		\hline  
		17	&TXPSC			 	 	& 2006-09-17 04:20			& A0-G0 		& 62.0 	& 73 		& 1.36 	& 1.140 \\
		\hline  
		18	&\textbf{TXPSC}\tablefootmark{a} 	& 2006-09-21 05:55			& K0-G0   		& 62.6 	& 72		& 1.39 	& 1.225 \\
			&HD48915				& 2006-09-21 08:36			& \ldots 		& \ldots 	& \ldots 	& 1.51	& 1.309 \\    
		\hline
		19	&TXPSC			 		& 2008-09-27 04:45	 		& G1-D0		& 63.6 	&133 	& 1.00 	& 1.155 \\ 
		\hline 
		20	& \textbf{TXPSC} 					& 2006-08-17 07:00	 		& A0-G0 		& 63.8 	& 73 		& 0.39 	& 1.137  \\
			&HD224935				& 2006-08-17 07:23			& \ldots 		& \ldots 	& \ldots	& 0.43	& 1.060 \\ 
		\hline                                    
		21	&\textbf{TXPSC}  					& 2004-10-30 03:28			& UT2-UT4	& 84.3 	& 80 		& 0.78 	& 1.244\\   
			&HD49161 				& 2004-10-30 09:19 			& \ldots 		& \ldots 	& \ldots 	& 0.72 	& 1.196 \\
		\hline
		22	&\textbf{TXPSC} 					& 2006-08-16 07:06			& A0-G1 		& 87.1 	&113 	& 0.85 	& 1.137 \\
			&HD18884				& 2006-08-16 07:57			& \ldots 		& \ldots 	& \ldots	& 0.81	& 1.353 \\ 
		\hline                                                        
		23	&TXPSC			 	 	& 2006-08-15 06:20			& A0-G1		& 90.3 	&113 	& 0.88 	& 1.145 \\
		\hline
		24	& TXPSC			 		& 2011-09-21 05:39			& I1-A1 		& 104.0 	& 84 		& 1.03 	& 1.186 \\
		\hline
		25	& TXPSC			 	 	& 2011-10-05 04:21 			& K0-A1		& 129.0 	& 67 		& 0.57 	& 1.158 \\
		\hline 
		26	&\textbf{TXPSC}\tablefootmark{a}		& 2011-10-02 05:44			& A1-J3		&140.0	& 46		& 1.02	& 1.305\\
			&HD45348				& 2011-10-02 06:42	 		& \ldots 		& \ldots 	& \ldots 	& 0.88	& 1.512 \\ 
		\hline                                                                                                         
		\end{tabular}
		\tablefoot{
		Calibrators that are used to calibrate the data are given below the science target.  The baseline 
		configuration, projected baseline length and position angle of the observations are given. The 
		observatory seeing and airmass give the average atmospheric conditions present at the observatory 
		during the observation. Observations that are used in the astrophysical interpretation are marked 
		in boldface.\\
		\tablefoottext{a}{Calibrated spectrum available}}
	\end{table*}
	}
    Data are reduced using MIA+EWS 1.7.1\footnote{\tt{http://www.strw.leidenuniv.nl/$\sim$jaffe/ews/MIA+EWS-\\Manual/index.html}} 
    \citep{jaffe04, ratzka05, leinert04}. 
    Observations are not used if one or more of the selection criteria discussed in \citet{klotz12a} are 
    violated (observations that are not boldfaced in Table\,\ref{journal}). 
    Uniform-disk angular diameters and IRAS 12\,$\mu$m flux of the calibrator targets are given in 
    Table~\ref{calibrators} (electronic version only). \\
   \onltab{2}{
	\begin{table}[]
		\begin{center}
		\begin{footnotesize}
		\caption{\label{calibrators}Properties of the calibrator targets.}
		\begin{tabular}{lllrr}
		\hline\hline
		HD & Name&Sp. T.\tablefootmark{a}&$F_{12}$\tablefootmark{a} & $\theta$\tablefootmark{b}\\
		 &&& [Jy] & [mas]\\
		\hline
		\object{HD 48915} & Sirius & A1 &143.1$\pm$3 & 6.08$\pm0.03$ \\
		\object{HD 20720} & $\tau$04 Eri & M3/M4 &162.7$\pm$6 &10.14$\pm$0.04 \\
		\object{HD 224935} & YY Psc& M3 &86.9$\pm$5 &7.25$\pm$0.03 \\
		\object{HD 49161} &17 Mon & K4 &10.4$\pm$5 &2.44$\pm$0.01 \\
		\object{HD 18884} &$\alpha$ Cet & M1.5 &234.7$\pm$3 & 12.28$\pm$0.05\\
		\object{HD 45348} &Canopus & F0 &154.8$\pm$3 & 6.87$\pm$0.03 \\
		\hline
		\end{tabular}
		\end{footnotesize}
		\end{center}
		\tablefoot{
		\tablefoottext{a}{\tt{http://www.eso.org/observing/dfo/quality/MIDI/qc/\\calibrators\_obs.html}}\\
		\tablefoottext{b}{\tt{http://simbad.u-strasbg.fr/simbad/}}\\
		}
	\end{table}
	}
    As only one suitable calibrator is available per observation, a standard multiplicative 
    error of 10\% is assumed for the calibrated visibilities \citep{chesneau07} . Some calibrated visibilities at baselines shorter than 
    $\sim$30\,m are significantly noisier and sometimes larger than unity after 11.5\,$\mu$m. Therefore, 
    in the following, spectro-interferometric observations are considered only for wavelengths shorter than 
    11.5\,$\mu$m.\\
    For the calibrated spectra additional selection criteria are applied \citep{chesneau07}: 
    (i) the airmass difference between science and calibrator observations is $<0.2$, 
    (ii) the spectral type of the calibrator is not later than M0. Considering these criteria, five spectra are derived  
    (flagged with `a' in Table\,\ref{journal}). \\
    TX\,Psc is almost unresolved for baselines shorter than 32\,m. Therefore, we 
    expect that the star is unresolved by the single-dish UTs. Thus, most of the mid-infrared (mid-IR) flux is located 
    within the field-of-view (FoV) of the UTs and consequently also in the FoV of the ATs and ISO, which 
    makes the spectra fully comparable to each other.

  \subsection{ISO spectra}
  \label{observations-iso}
    Three spectra of TX\,Psc were observed with the Short-Wavelength-Spectrometer \citep[SWS,][]{degraauw96} onboard 
    of ISO \citep{sloan03,jorgensen00}. Two spectra have a resolution of $\mathrm{R}\sim200$ and range from 
    $2.36-45.35$\,$\mu$m. The other spectrum ranges from $2.45-45.20$\,$\mu$m and has a resolution of 
    $\mathrm{R}\sim2\,000$ which has been binned to the resolution of the other spectra. 
    For the ISO spectra a multiplicative error of $\pm10$\% is assumed from 2.38-4.05\,$\mu$m and $\pm5$\% 
    afterwards \citep{sloan03}. 
    
    \subsection{Interferometric and spectroscopic variability}    
    \label{variability}
\begin{figure*}
		\centering
		\includegraphics*[width=8.9cm,bb=80 370 555 705]{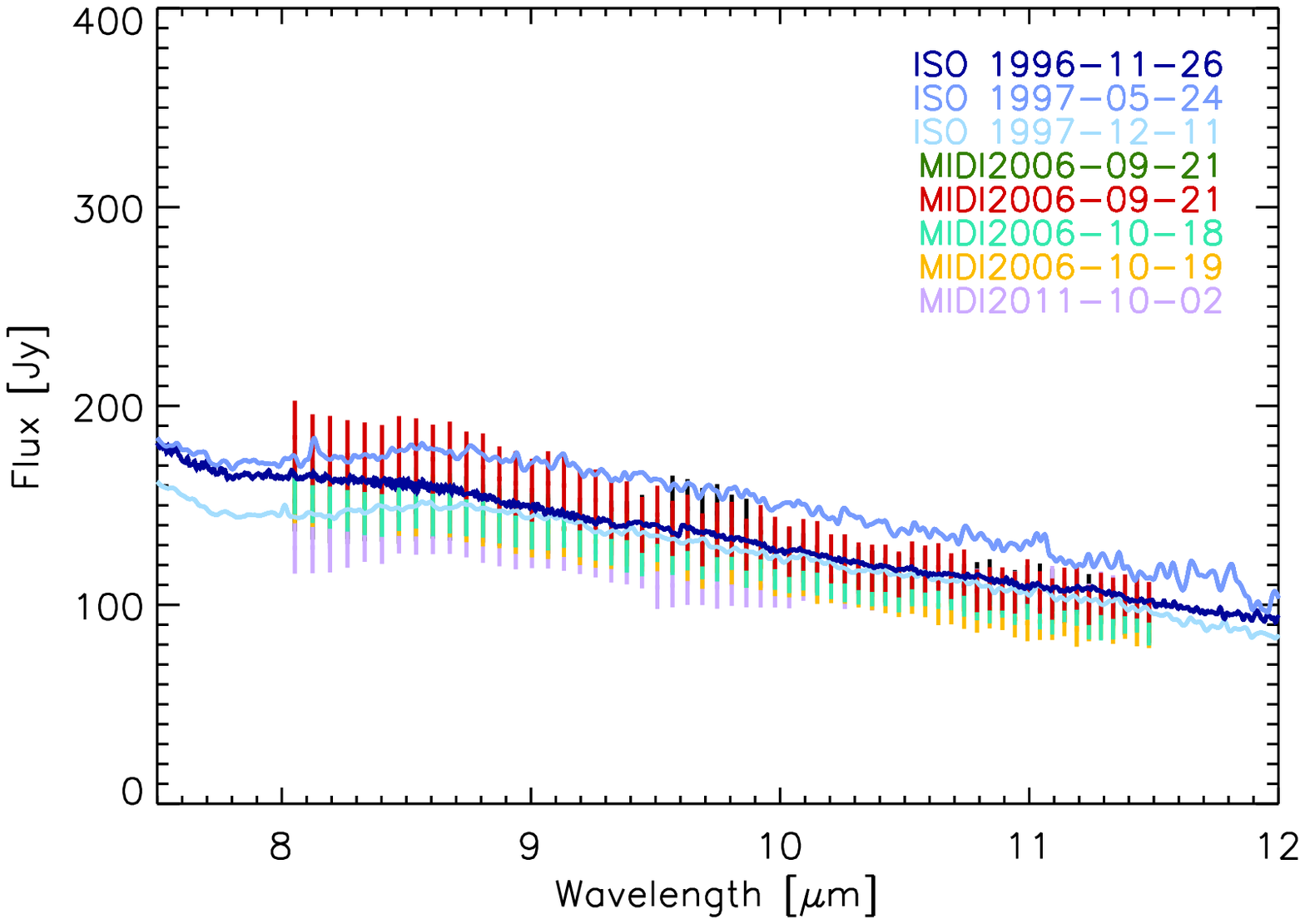}
        \includegraphics*[width=8.9cm,bb=80 370 555 705]{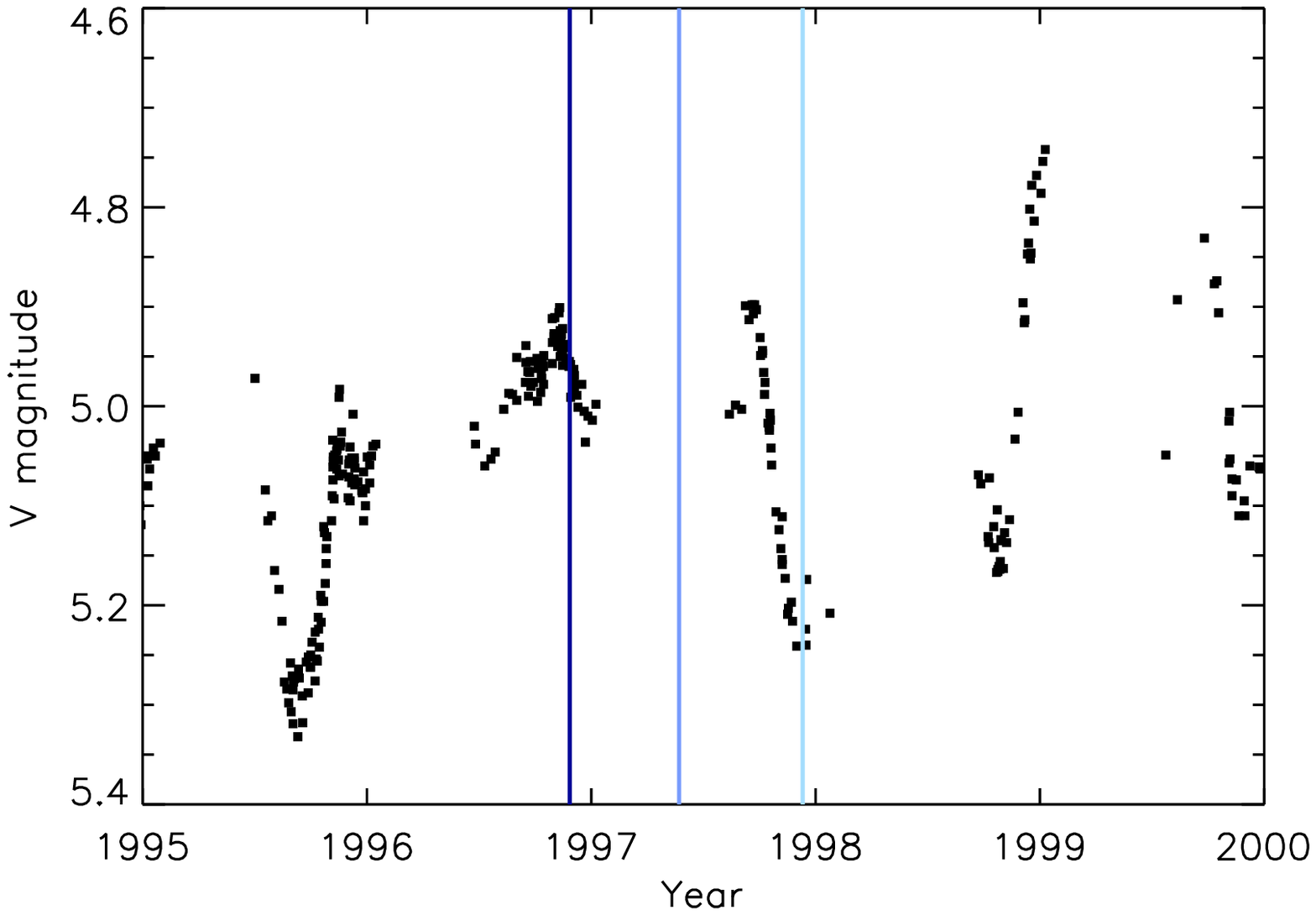}
		\caption{\label{iso} Left: MIDI flux (error bars) for different dates. Overplotted are the three ISO/SWS spectra 
		(full lines) of TX\,Psc. Right: AAVSO $V$-band light curve of TX\,Psc. Vertical lines mark the observation date of 
		the ISO spectra. Colors are the same as in the left figure. }
	\end{figure*}

    In the left panel of 
    Fig.~\ref{iso} the calibrated MIDI spectra are overplotted to the ISO/SWS spectra.  The flux level of the MIDI 
    observations is the same (within the error bars) as the ISO/SWS spectra taken at 1996 Nov 26 and 1997 Dec 11.
    This suggests that no significant cycle-to-cycle variation is expected in the mid-IR. 
    The right plot of Fig.~\ref{iso} shows part of the lightcurve of TX\,Psc in $V$ taken from AAVSO. The spectra 
    from 1996 Nov 26/1997 Dec 11 were observed at a 'local' visual maximum/minimum, respectively.
    In the following we will use these two spectra to derive the parameters 
    of the star at different phases. As we do not have any information on the $V$ magnitude of the spectrum from 1997 
    May 24, this spectrum is not used in the subsequent sections. Additionally, the phase for all the MIDI observations 
    is unknown. However, AAVSO visual estimates from one specific observer revealed that none of the MIDI 
    observations were taken at visual minima or maxima.\\
    Simulations with dust-free model atmospheres show that the effect of pulsation on 
    the $N$-band UD-radius is smaller than 0.1 mas at 1 kpc \citep{paladini09}. For a star at the distance of TX~Psc this 
    would be $\sim$0.3 mas. This value is beyond the limit of resolution of MIDI. Therefore, we do not expect any 
    observed intra-cycle variability effect that is larger than the errors. This allows to combine all calibrated 
    visibilities for the further analysis. 
	%


\section{Model description}
\label{model-description}
In the following we present the different classes of models and the derivation of synthetic 
observables that will be compared to observations in Sect.\,\ref{parameterdetermination}.

  \subsection{Hydrostatic models}
  \label{mod.sect}
    Observed spectra and visibilities are compared to the grid of spherical hydrostatic model atmospheres and spectra of 
    \citet{aringer09}. These models are computed with COMARCS and are generated assuming hydrostatic local thermal and chemical equilibrium.  
    The molecular and atomic opacities are treated in the opacity sampling (OS) approximation. The parameters that 
    characterize a model are: effective temperature T$_\mathrm{eff}$, metallicity $Z$, surface gravity $g$, mass $M$, 
    and carbon to oxygen ratio C/O.   \\ 
      For this work we limit the sample to  
      models having solar metallicity as the effect of metallicity is expected to be small for low-resolution 
      spectroscopy. Additionally, there is no indication for largely non-solar metallicity from other properties 
      of this star. \\
      The spectra cover the following parameters: $2\,400\le T_{\rm{eff}}\le4\,000$ K with steps of 
      100 K; $Z/Z_{\sun} = 1$; $-1.0\le $log$(g[$cm\,s$^{-2}]) \le +0.0$; $M/M_\odot = 1, 2$; C/O$ = 1.05, 1.10, 
      1.40, 2.00$. In order to get precise estimates of T$_\mathrm{eff}$, additional model atmospheres were 
      produced resulting in a grid spacing of $\Delta$T=10\,K. All the main molecular 
      opacities typical for C-stars were included: CO \citep{goorvitch94}, C$_2$ 
      \citep{querci74}, HCN \citep{harris06}, CN \citep{jorgensen97} in the form of line lists, while C$_2$H$_2$ and 
      C$_3$ \citep{jorgensen89} as OS data. CS is not included due to the lack of line lists and OS data. 
      Synthetic spectra with a resolution of 18\,000 are 
      computed in the wavelength range $0.8-25\,\mu$m. The spectra are convolved in order to get the same 
      resolution as the observed data.\\
      Among the output of the spherical radiative transfer code COMA is the monochromatic spatial intensity profile. 
     This profile is used to calculate a synthetic visibility profile in the mid-IR for a subset of the models in the 
     grid. A detailed 
     description of the computation of the visibility profiles is given in \citet{paladini09}. 

  \subsection{Evolutionary tracks}
    The luminosity and effective temperature that are determined from hydrostatic models are compared to 
    thermally-pulsing (TP) AGB evolutionary tracks from Marigo et al. (in prep). We selected TP-AGB sequences with 
    an initial scaled-solar chemical composition ($Z=0.014$, $Y=0.273$), where $Z$ and $Y$ denote the mass 
    fractions of metals and helium, respectively. TP-AGB evolutionary calculations are carried out from the first 
    thermal pulse - extracted from the PARSEC database of stellar models \citep{bressan12} - to the complete 
    ejection of the envelope due to stellar winds. The TP-AGB tracks are based on numerical integrations of complete 
    envelope models in which, for the first time, molecular chemistry and gas opacities are computed on-the-fly with 
    the \AE SOPUS code \citep{marigo09}. This guarantees a full consistency of the envelope structure with the surface 
    chemical abundances that may significantly vary due to the third dredge-up episodes and hot-bottom burning. 
    The transition from C/O$~<1$ to C/O$~>1$ is followed accurately, in particular in the narrow range from 
    $0.95\approx$ C/O $\approx 1.05$, where an abrupt change in the molecular chemistry and opacity is expected 
    to occur \citep[see figures 11 and 16 in][]{marigo09}. This point is particularly relevant in the context of the present 
    work, as TX Psc is found to have a surface C/O slightly above unity.
  
 
\section{Stellar parameter determination}  
\label{parameterdetermination}
  Stellar parameters for TX\,Psc were determined by a number of authors. A summary is given in Table\,\ref{parameters}. \\
  \begin{table*}[]
	\begin{center}
	\begin{footnotesize}
	\caption{\label{parameters}Published stellar parameters of TX\,Psc.}
	\begin{tabular}{lccccccc}
	\hline\hline
	Reference &  $T_{\mathrm{eff}}$ & log\,$g$ & Mass &  C/O  & $\theta$ & $\lambda_\theta$& d  \\
	& [K] &&[$M_\odot$] &  & [mas] &[$\mu$m]&[pc]\\
	\hline\\[-5pt]
	\citet{lasker73}&  && & &9.00 &0.66 \\
	\citet{devegt74} & &&  & &8.00 &0.71\\
	\citet{dunham75} & &&  & &10.20 &0.69 \\
	\citet{lambert86} & 3\,030 & 0.0&  & 1.03 & \\
	\citet{claussen87} & && & & & &280\\
	\citet{quirrenbach94} & 2\,805 &  && & 11.20 &0.7-0.8  \\
	\citet{richichi95} &  && & & 8.38 &0.55-3.60  \\
	\citet{dyck96} & 2\,921&  && & 11.20 &2.2 &\\
	\citet{jorgensen00} & 3\,000 & -0.5 & & 1.02 &  \\
	\citet{ohnaka00} & 3\,000 &&  & 1.07 &  \\
	& 3\,100 &&  & 1.17 &  \\
	\citet{bergeat01} & 3\,115 &&  & & \\
	\citet{harris03} & 3\,050 & 0.0&  & 1.02 &  \\
	\citet{gautschy04} & 3\,200 &-0.3&1 &1.10 &  \\
	\citet{bergeat05} & 3\,125  && & & && 315 \\
	\citet{ragland06} & && &  &9.89 &1.65 \\
	\citet{vanleeuwen07} & & & & && &275$^{+34}_{-26}$\\\\[-5pt]
	\hline
	\end{tabular}
	\end{footnotesize}
	\end{center}
  \end{table*} 
  In the following two different approaches will be used to determine the stellar parameters of TX\,Psc: 
  (i) in Sect.\,\ref{methoda} we use the `classical approach' of interferometrists where spectro-interferometric observations 
  are used to  determine the effective temperature $T_{\mathrm{eff}}$; 
  (ii) in Sect.\,\ref{methodb} we follow the approach of \citet[P11 hereafter]{paladini11} where spectroscopic/interferometric 
  measurements are compared to hydrostatic models to determine $T_{\mathrm{eff}}$ and C/O ratio and to constrain mass and log\,$g$.  

  \subsection{Approach 1: Geometric models}
  \label{methoda}    \label{ang-diam}
\begin{figure*}
	\centering
	\includegraphics*[width=8.9cm,bb=80 370 555 705]{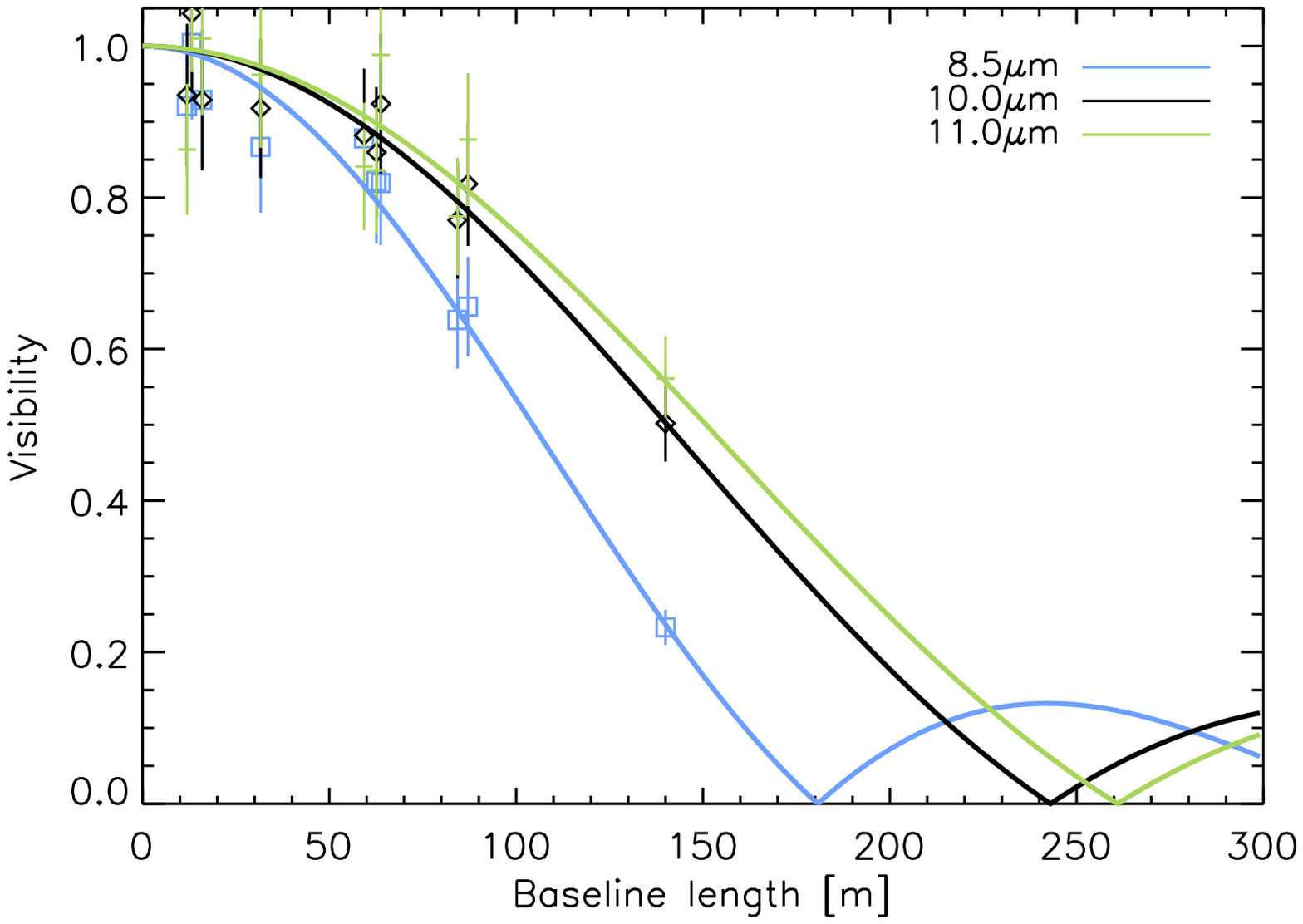}
	\includegraphics*[width=8.9cm,bb=80 370 555 705]{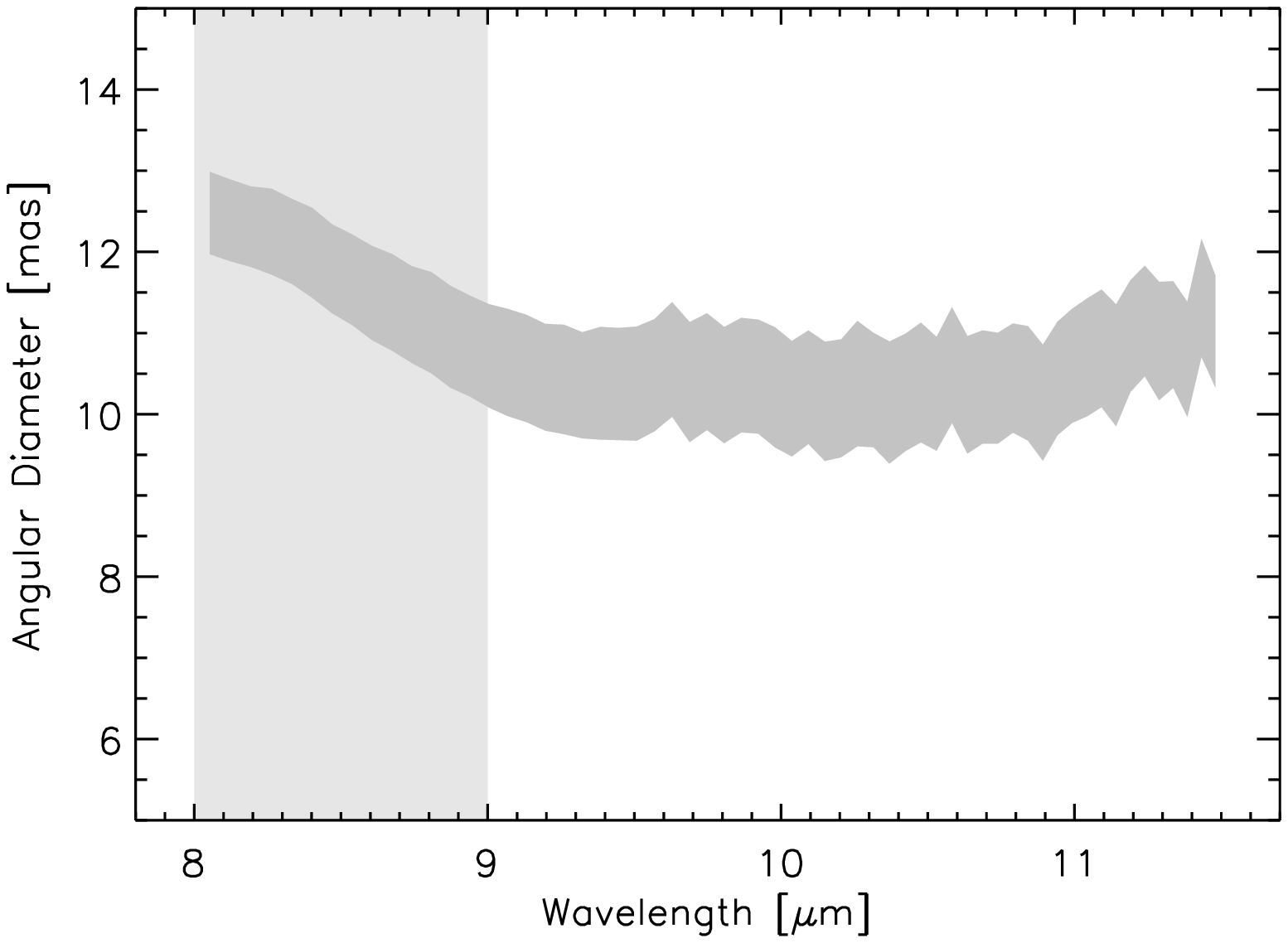}
	\caption{\label{diameter} Left: Calibrated visibilities (symbols) versus baseline length for three different 
	wavelengths. The lines represent the best-fitting UD-model at the given wavelength. Right: Spectrally dispersed 
	angular diameter plus errors from the best-fitting UD-model (dark grey shaded area).  
	The light grey shaded area marks the region that is omitted for the mean diameter estimation.
	}
  \end{figure*}

    Interferometry is a powerful tool for constraining the morphology and size 
    of stars and their environments. \\
    A deviation from spherical symmetry for the circumstellar environment of TX\,Psc was detected 
    by several authors at different spatial scales and wavelengths 
    \citep[e.g.][Hron et al in prep.]{cruzalebes98,ragland06,sacuto11b,jorissen11}. 
    Clearly, the circumstellar environment of TX\,Psc is very complex and most likely unrelated effects shape the 
    morphology in different regions, resulting in a large variety of structures. To study the geometry of the innermost 
    region of the star in the $N$-band, the geometrical model fitting tool GEM-FIND \citep{klotz12b} was used to fit 
    the MIDI interferometric observations of TX\,Psc. 
    A spherical UD-model is able to reproduce the calibrated visibilities, i.e.\,TX\,Psc can be assumed to be 
    spherically symmetric in the mid-IR at spatial scales probed by 
    our MIDI observations. The reason why the asymmetrical structures detected by other works are not observed is that 
    they were either detected at 
    distances outside the FoV of MIDI or at smaller spatial scales beyond 
    the detection limit of MIDI. This is supported by the differential phase measurements 
    of MIDI, which do not show any deviation from zero. The calibrated visibilities are plotted together with 
    the best-fitting UD-model in the 
    left panel of Fig.\,\ref{diameter} for three different wavelenghts.  \\
      The right panel  of Fig.\,\ref{diameter} shows the wavelength dispersed diameter calculated with the UD-model 
      (dark grey shaded area). The star appears larger between 8 and 9\,$\mu$m. According to \citet{gautschy04} 
      CS is affecting this wavelength range. Because of this molecular contamination, this region is omitted 
      and a mean angular diameter of $\theta=10.51\pm0.70$\,mas 
      is calculated by averaging the diameter from 9-11.5\,$\mu$m. 

    \subsubsection{Effective temperature}
    \label{model-indep-teff}
\begin{figure}
		\centering
		\includegraphics*[width=8.9cm,bb=80 365 555 705]{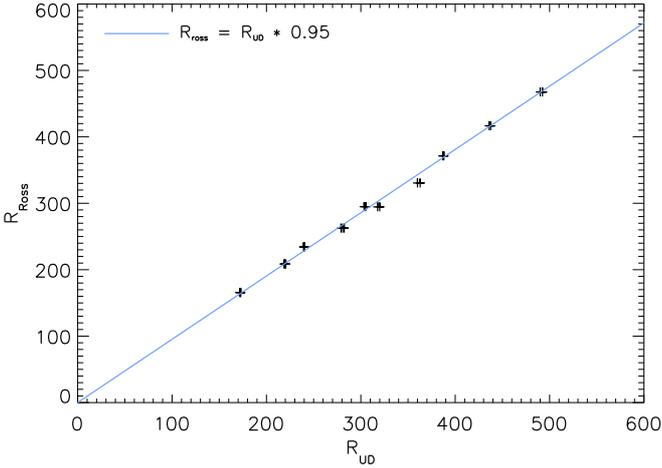}
		\caption{\label{rross-vs-ud} Rosseland radius of the synthetic models versus the derived mid-IR UD-radius of the same models 
		(crosses). The blue line is a linear fit to the data.}
	  \end{figure}

      The temperature can be determined using the apparent bolometric magnitude 
      $m_\mathrm{bol}$ and the angular Rosseland diameter $\theta_\mathrm{ross}$.
     The apparent bolometric magnitude $m_\mathrm{bol}=2.26$\,mag was derived by F.\,Kerschbaum 
     by fitting a combination of blackbodies to near-IR and IRAS data\footnote{Method described in
     \citet{kerschbaum96} and references therein; near-IR data from the IRAS catalogue and \citet{fouque92}.}.\\
     Various definitions for the radius can be found in literature \citep[c.f. reviews by][]{baschek91,scholz03}, where the most 
     commonly used radius in atmospheric modeling is the Rosseland radius. It is defined by the distance between the 
     center of the star and the layer having Rosseland optical depth $\tau_{\mathrm{ross}}=\frac{2}{3}$. This radius, however, 
     is not an observable quantity and observed radii have to be converted 
     by using model considerations. 
     In the following we will derive this conversion factor for hydrostatic C-stars by using a subset 
     of the hydrostatic models in the grid of \citet{aringer09} to derive a mean 
      UD-radius in the mid-IR ($9-11.5$\,$\mu$m). 
      This mean UD-radius is plotted versus the Rosseland radius of the corresponding hydrostatic model in Fig.\,\ref{rross-vs-ud}.
      There is a clear correlation between the two radii, yielding
      \begin{equation}
          R_\mathrm{Ross}=0.95\;R_\mathrm{UD}\;.
      \end{equation}   
      This implies that for a hydrostatic C-star the Rosseland radius can be approximated by 
      the mid-IR UD-radius if the derived correction factor of 0.95 is applied.  \\
     Applying this correction factor yields an angular Rosseland diameter $\theta_\mathrm{Ross}=9.99$\,mas. Together with the bolometric 
     magnitude the distance-independent effective temperature can be derived. 
     Using the three different distance estimates $d$ that are available for TX\,Psc (see Table\,\ref{parameters}) a linear radius and 
     luminosity can be calculated. 
      Table\,\ref{determinedparameters-a1} lists the derived and calculated stellar parameters of approach 1. Errors on the temperature 
      are determined by using 
     the errors on $\theta_\mathrm{UD}$ and by assuming an arbitrary error of $\pm0.1$ for m$_\mathrm{bol}$ that accounts for the 
     stellar variability and the fitting error. These values are in agreement with 
     those given in literature (see Table\,\ref{parameters}).
     \begin{table}[]
		\begin{center}
		\begin{footnotesize}
		\caption{\label{determinedparameters-a1} Stellar parameters derived from observations (middle block) and 
		calculated (right block) using approach 1.
		}
		\begin{tabular}{l|l|llll}
		\hline\hline
		$d$  &$\theta_{UD}$ & $\theta_{Ross}$ & $T_\mathrm{eff} $ & $R$  & $L $  \\
		$[$pc] &  [mas] &  [mas] & [K]  & [$R_\odot$] & [$L_\odot$]\\
		\hline
		&&&&\\
		275 &  10.51$\pm$0.70 & 9.99& 3127$^{+192}_{-173}$ & 294 & 7\,406\\
		 &&&&\\
		 280 & 10.51$\pm$0.70 & 9.99 &3127$^{+192}_{-173}$ & 299 & 7\,678\\
		 &&&&\\
		315  &  10.51$\pm$0.70 & 9.99 &3127$^{+192}_{-173}$ & 337 & 9\,717\\
		&&&&\\
		\hline
		\hline
		\end{tabular}
		\end{footnotesize}
		\end{center}
	  \end{table} 	  

  \subsection{Approach 2: Hydrostatic models}
  \label{methodb}
    The short wavelength part of the ISO spectrum of TX\,Psc is dominated by the 3\,$\mu$m feature which has 
    contributions from HCN and C$_2$H$_2$. 
    The 5\,$\mu$m feature, on the other hand, is due to C$_3$ and the fundamental band of CO. The region from 7 to 
    8\,$\mu$m is dominated by HCN and C$_2$H$_2$ and according to \citet{gautschy04} the region from 8 to 9\,$\mu$m 
    is affected by CS opacity. The low variability in the $V$-band 
    ($\Delta V\sim0.4$\,mag) justifies the use of 
    hydrostatic models in the near- as well as mid-IR. 
    Additionally, hydrostatic models are able to reproduce large parts of the overall ISO 
    as well as MIDI spectra. This indicates that the circumstellar environment of TX\,Psc 
    is optically thin and contains almost no dust.\\
    In Sect.\,\ref{co-hydro} and \ref{temp-hydro} low-resolution spectroscopic observations are compared with synthetic 
    spectra of hydrostatic models to fix the fundamental stellar parameters C/O ratio and  $T_{\mathrm{eff}}$. The 
    overall energy distribution as well as the bands of the molecules that are present from 2.3-6\,$\mu$m put strong 
    constraints on these parameters \bibpunct[; ]{(}{)}{;}{a}{}{;}\citep[][P11]{jorgensen00,loidl01}.\bibpunct[, ]{(}{)}{;}{a}{}{;}\\
    Low resolution spectroscopy does not allow to ascertain mass and log\,$g$ (Fig.\,6-10 in P11). To determine 
    these parameters Sect.\,\ref{masslogg} follows the approach described in P11: spectro-interferometric observations are 
    compared to models of fixed  $T_{\mathrm{eff}}$ and C/O ratio but varying log\,$g$ and mass.

    \subsubsection{C/O ratio}
    \label{co-hydro}
\begin{figure*}
		\centering
		\includegraphics*[width=\textwidth,bb= 73 375 509 566]{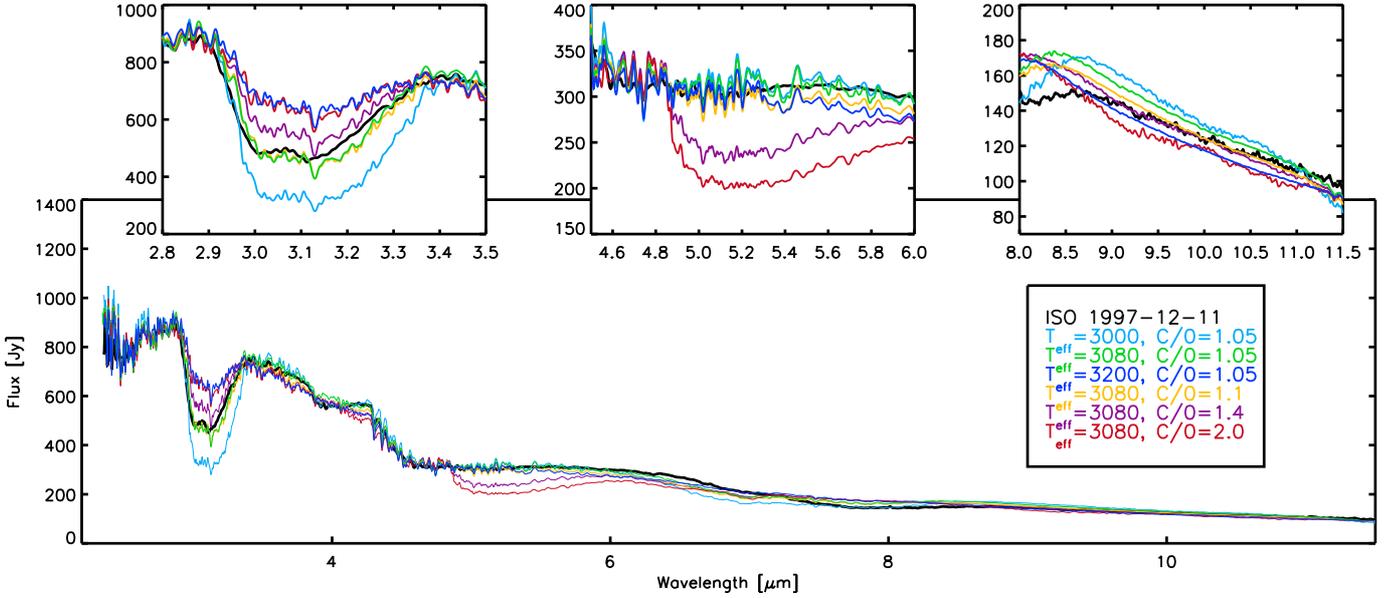}
		\caption{\label{bestfittingmodel} ISO/SWS spectrum of TX\,Psc at visual minimum from 1997\,Dec\,11 (black line) plotted with 
		hydrostatic models (colored lines) of different temperature and C/O ratio. The best fitting model is plotted in green.
		Model spectra are normalized to the flux of the 
		corresponding ISO spectrum at 2.9\,$\mu$m.}
	  \end{figure*}

\begin{figure*}
		\centering
		\includegraphics*[width=\textwidth,bb= 73 375 509 566]{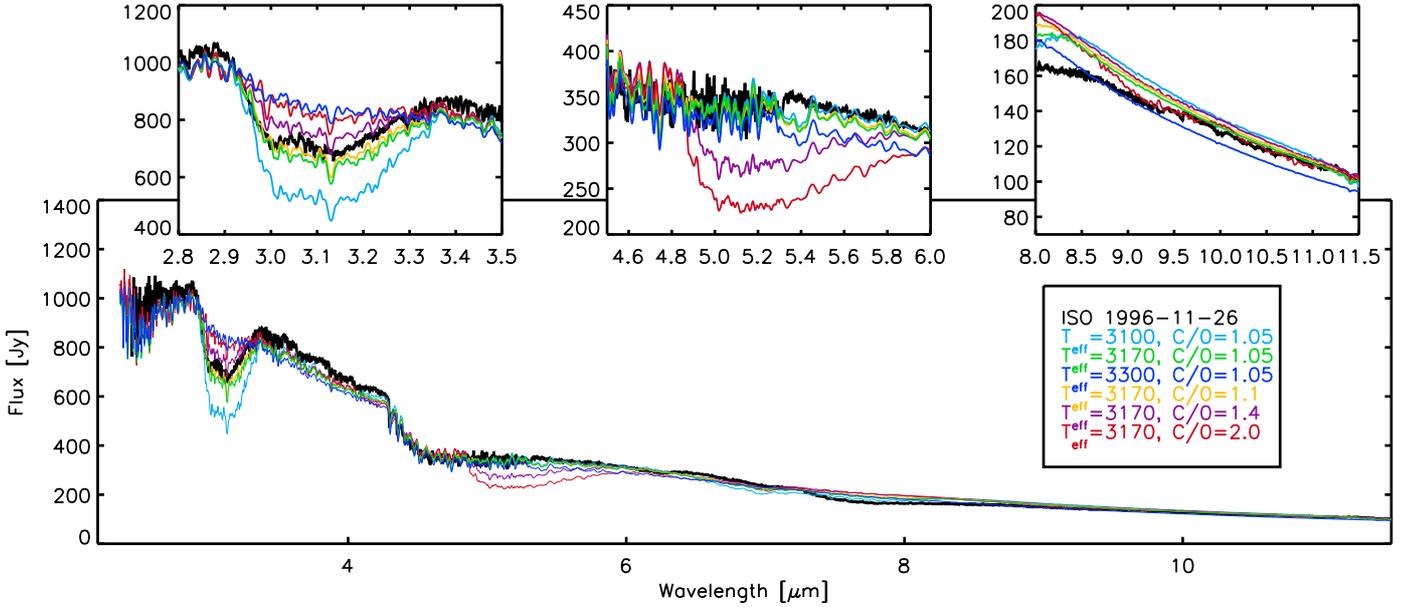}
		\caption{\label{bestfittingmodel2} Same as Fig.\,\ref{bestfittingmodel}, but for at visual maximum from 1996\,Nov\,21.}
	  \end{figure*}

      \citet{jorgensen00} found the ratio between the 3\,$\mu$m feature (HCN and C$_2$H$_2$) and the 5.1\,$\mu$m 
      feature (CO and C$_3$) to be a sensitive measure of the C/O ratio. In order to be independent of distance each model 
      spectrum is normalized to the ISO flux at 2.9\,$\mu$m \citep[local minimum of molecular absorption,][]{aringer09}. 
      A $\chi^2$ test is applied between 2.9 - 6.0\,$\mu$m 
      to compare the observed ISO spectra 
      (1996 Nov 26 and 1997 Dec 11) to the models. For both ISO spectra 
      we find that the best solution is obtained with a C/O ratio of 1.05. Considering all solutions lying within the 68\% 
      confidence level a C/O ratio of 1.1 can be defined as an upper limit. Due to the coarse grid spacing for C/O this value is 
      not a strict upper limit and no lower limit can be assigned. Figs.\,\ref{bestfittingmodel} and \ref{bestfittingmodel2} 
      show the ISO/SWS spectrum plotted together with models of different temperatures and C/O ratios. The upper panels draw 
      the region around the 3\,$\mu$m and 5.1\,$\mu$m feature to a larger scale. These plots demonstrate that a synthetic 
      spectrum with a C/O ratio larger or equal 1.4 is not able to reproduce the observations, because it over-evaluates the depth 
      of the 5.1\,$\mu$m feature. This finding is consistent with the C/O ratios given in literature (see Table\,\ref{parameters}).

    \subsubsection{Effective temperature}
    \label{temp-hydro}
      We expect the 3\,$\mu$m feature to be a good temperature indicator for hydrostatic stars (P11). Consequently, in order to find the best 
      temperature for TX\,Psc, a $\chi^2$ test is used to compare the observed ISO spectra and the model spectra between 2.9 - 
      3.6\,$\mu$m. Only model spectra lying within the confidence level of C/O (1.05, 1.1; see Sect.\,\ref{co-hydro}) are used 
      for this test. \\
      The large grid of models allows to determine T$_\mathrm{eff}$ very precisely. All solutions lying within the 68\% confidence
      level have a temperature of 3\,080$^{+70}_{-60}$ / 3\,170$^{+70}_{-80}$\,K for the visual minimum/maximum, respectively. The upper 
      left panel of Figs.\,\ref{bestfittingmodel} and \ref{bestfittingmodel2} show that models with higher/lower temperatures do not reproduce the depth
      of the 3\,$\mu$m feature. \\
      The temperature derived with approach 1 in Sect.\,\ref{model-indep-teff} is within the errors of the temperature that is 
      derived here for the visual minimum and maximum. \\
      We confirm the finding of \citet{jorgensen00} that the difference in the two ISO spectra can be explained by a temperature 
      change. The lightcurve suggests that this difference is due to variability effects, but time-series spectroscopy is needed to 
      confirm this finding.
 
    \subsubsection{Surface gravity and mass}
    \label{masslogg}
    \label{diam-hydro}
\begin{figure*}
		\centering
		\includegraphics*[width=\textwidth,bb=65 523 549 717]{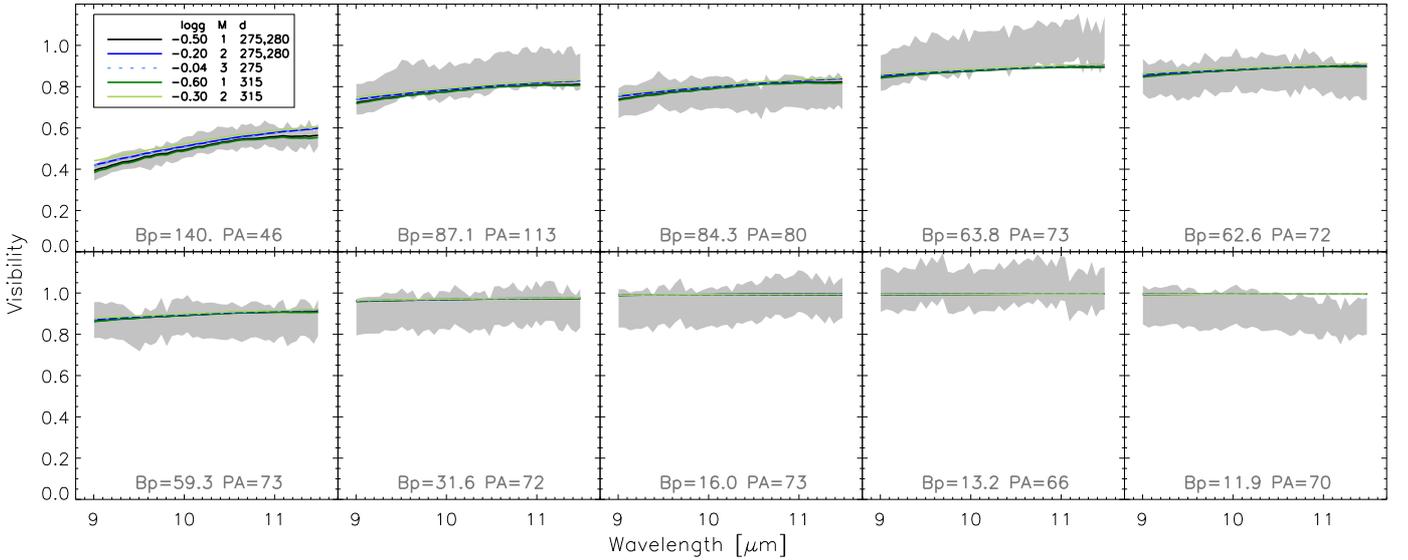}
		\caption{\label{logg-mass} Wavelength-dispersed calibrated visibilities plus errors (dark grey shaded area) plotted 
		with the best-fitting synthetic visibilities of hydrostatic models (full lines). To show the 
		degeneracy between mass and log\,g one model with 3\,M$_\odot$ is overplotted (dotted line).  }
	 \end{figure*}

      Interferometric observations are compared to synthetic visibilities of hydrostatic models of varying log\,g and mass 
      using a $\chi^2$ test.
     All synthetic visibilities are computed for the best-fitting values of T$_\mathrm{eff}$ and C/O-ratio from 
     Sects.\,\ref{co-hydro} and \ref{temp-hydro}. Synthetic visibilities are derived at
      3 different distances: 275\,pc, 280\,pc, 315\,pc 
      \citep[][respectively]{vanleeuwen07,claussen87,bergeat05}. In Fig.\,\ref{logg-mass} the wavelength-dispersed calibrated visibilities are plotted together with the 
      synthetic visibilities of the best-fitting models. The region between 8 - 9\,$\mu$m is not considered in the fitting, as the 
      hydrostatic models do not include the CS opacity data. \\
      The middle block of Table\,\ref{determinedparameters} gives the best-fitting stellar parameters for a given distance that were determined using 
      approach 2. 
      It is clear from the table and from Fig.\,\ref{logg-mass} that, given the error bars on the visibilities, there 
      is a degeneracy between log\,$g$ and mass. 
      As the distance defines the level of visibility, there is also a degeneracy between distance and log\,$g$. 
      From current TP-AGB evolutionary calculations \citep[e.g.][Marigo et 
      al.\,in prep.]{karakas02,marigo07} we expect that a $1\,M_{\odot}$ TP-AGB star with solar metallicity does not make the 
      transition to the C-rich domain. 
     This suggests that the log\,$g$ values in Table\,\ref{determinedparameters} found for $M=2$\,$M_\odot$ are the more reliable ones.      
      But, considering the degeneracy 
      and the limited mass sampling ($M=1,2$\,$M_\odot$) in the grid, also models with higher masses 
      would reproduce the observed visibilities. To support this statement, one additional model with 3\,$M_\odot$ is calculated and 
      overplotted in Fig.\,\ref{logg-mass} as dashed line. The models are almost indistinguishable.\\
      The right part of Table\,\ref{determinedparameters} gives stellar parameters that are calculated from the derived log\,$g$, mass 
      and effective temperature. 
	  The luminosities are significantly larger than the $L=5\,200$\,$L_\odot$ used by \citet{gautschy04}, but 
      comparable to $L=7\,700$\,$L_\odot$ derived by \citet{claussen87}. The luminosities and radii are also in perfect agreement 
      with the ones determined with approach 1 (see Sect.\,\ref{methoda}). 
	  
	  \subsubsection{Photometric constraints on the best model}
\begin{figure}
		\centering
		\includegraphics*[width=9.cm]{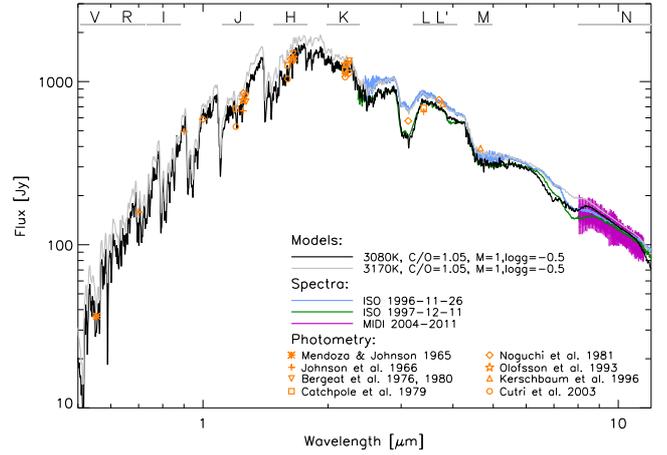}
		\caption{\label{sed-plot}   Best fitting hydrostatic models for the visual minimum (black line) and maximum (grey line). 
		Superimposed are the ISO spectra  of visual minimum (green line) and visual maximum (blue line), MIDI spectra (violet lines) 
		and photometric measurements of different works (orange symbols). Model spectra are normalized to the flux of the 
		corresponding ISO spectrum at 2.9\,$\mu$m.}
	  \end{figure}

      The best fitting hydrostatic models of visual minimum/maximum (model with lowest $\chi^2_\mathrm{logg,M}$ in 
      Table\,\ref{determinedparameters}) are overplotted to the ISO and MIDI spectra as well as to photometric 
      measurements in Fig.\,\ref{sed-plot}. Photometric measurements from \citet{johnson66}, \citet{mendoza65} 
      and \citet{catchpole79} were observed with the Johnson filter system. Zero points to convert these 
      measurements from magnitudes to Jansky are taken from \citet{allen00}. These zero points are also 
      used to convert observations from \citet{bergeat76} and \citet{bergeat80} as the authors claim that their filter 
      system is similar to the Johnson filter system. Zero points for 2MASS photometry \citep{cutri03} are given in 
      \citet{cohen03}. \citet{olofsson93} and \citet{kerschbaum96b} used the ESO filter system. Zero points are 
      taken from \citet{lebertre88} and \citet{wamsteker81}. \citet{noguchi81} used their own filter system and 
      corresponding zero points are given in their paper.
	  \begin{table*}[]
		\begin{center}
		\begin{footnotesize}
		\caption{\label{determinedparameters} Stellar parameters derived from observations (middle block) and 
		calculated (right block) using approach 2.
		}
		\begin{tabular}{l|llllll|lll}
		\hline\hline
		$d$  & $T_\mathrm{eff,min} $ & $T_\mathrm{eff,max} $ &C/O &  log\,$g$ & $M$ & $\chi^2_\mathrm{logg,M}$ & $R$ & $L_\mathrm{min} $ & $L_\mathrm{max} $\\
		$[$pc] & [K] & [K] &&& [$M_\odot$]  & & [$R_\odot$]  & [$L_\odot$]  & [$L_\odot$] \\
		\hline
		&&&&&&&&\\
		275 & 3080$^{+70}_{-60}$ & 3170$^{+70}_{-80}$ & 1.05 & -0.5 & 1 & 0.50 & 295 & 7019 &7876 \\
		 &  &  &  & -0.2 & 2 & 0.50 & 295 & 7019 & 7876 \\
		 &&&&&&&&\\
		 280 & 3080$^{+70}_{-60}$ & 3170$^{+70}_{-80}$ & 1.05 & -0.5 & 1 & 0.48 & 295 & 7019 &7876 \\
		 &  &  &  & -0.2 & 2 & 0.53 & 295 & 7019 & 7876 \\
		 &&&&&&&&\\
		315        &  3080$^{+70}_{-60}$ & 3170$^{+70}_{-80}$ & 1.05 & -0.6 & 1 & 0.49 & 331 & 8836 & 9915\\
		 &  &  &  & -0.3 & 2 & 0.53 & 331 & 8836 & 9915\\
		&&&&&&&&\\
		\hline
		\hline
		\end{tabular}
		\end{footnotesize}
		\end{center}
	  \end{table*} 	 
      %


\section{Comparison with evolutionary tracks}
\label{evoltracks}
\begin{figure*}
	\centering
	\includegraphics*[width=14.95cm]{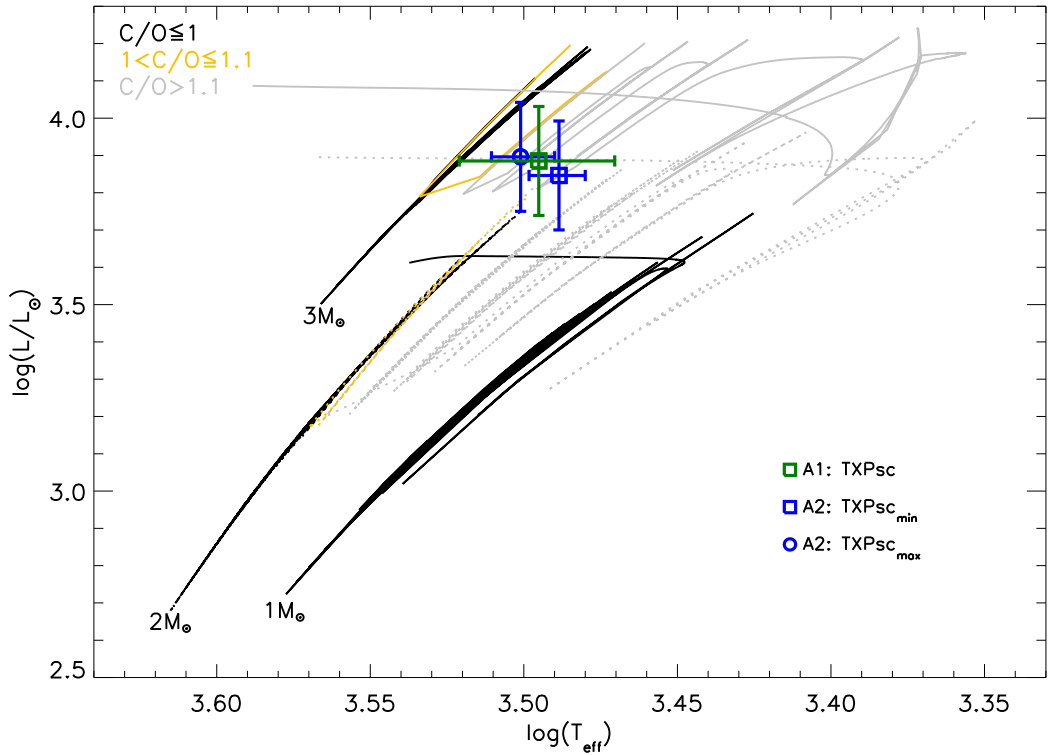}
	\caption{\label{evolutionary-tracks} Zoom into the AGB region of the H-R diagram.  Lines denote solar metallicity 
	evolutionary tracks (Marigo et al., in prep.) and numbers indicate the mass on the early-AGB. Yellow/grey lines mark 
	the region of carbon-rich AGB stars with C/O$>$1.0. Black lines mark the region of oxygen-rich AGB stars (C/O$\le$1). For better 
	visibility, the track of the 2 \,$M_\odot$ model is plotted with a dotted line. Different colored symbols refer to the 
	luminosity and effective temperature determined in this work (for the two different approaches A1 and A2 at visual minimum/maximum).}
  \end{figure*}

  We follow the approach described in P11 and compare the stellar parameters with new 
  evolutionary tracks of thermally pulsing AGB stars (Marigo et al., in prep). \\
  Figure\,\ref{evolutionary-tracks} depicts evolutionary tracks in the region of AGB stars in the H-R diagram. 
  Overplotted are the determined luminosity and temperature for
  TX\,Psc for the two approaches.
   Only the best-fitting luminosity (see Table 
  \ref{determinedparameters}) at d=280\,pc is plotted for TX\,Psc. Errors for the luminosity are assumed to lie in the order of $\sim$40\% 
  (uncertainty on the given distance measurement). \\
  We note an encouraging agreement between the observed location of TX Psc in the H-R 
  diagram and the predicted ranges of luminosity and effective temperature for a carbon-rich star with 
  solar-metallicity and C/O within a narrow interval (i.e. 1$<$C/O$\le$1.1). As we see in 
  Fig.~\ref{evolutionary-tracks}, the TP-AGB evolutionary tracks in the C-rich regime 
   extend to much 
  lower effective temperatures than the derived values for TX Psc. This cooling is mainly driven by the increase of the 
  C/O ratio after each dredge-up episode, as well as by the progressive strengthening of the mass-loss efficiency. 
  The relatively warm effective temperature of TX Psc suggests that this star is observed close after the transition into 
  the C-star domain, in an early `quiet' stage in which the strong wind has not yet developed. This 
  picture is nicely supported by the observational findings already discussed in the previous sections.\\
  It is visible from the evolutionary tracks that an AGB star with solar metallicity will turn into a carbon-rich AGB star 
  only for masses around 2\,$M_\odot$ and higher. Also, the position of TX\,Psc in Fig.\,\ref{evolutionary-tracks} 
  suggests that the mass lies between 2 and 3\,$M_\odot$. This is not in agreement with the best fitting models 
  having 1\,$M_\odot$ (Sect.\,\ref{masslogg}), but in good agreement with the models having 2\,$M_\odot$.


\section{Conclusion}
\label{conclusion}
  In this work we determined stellar parameters for TX\,Psc by comparing observations to geometric models \citep{klotz12b}, 
  state-of-the-art hydrostatic model atmospheres \citep{aringer09} and evolutionary models (Marigo et al.\, in prep.).  
  Two different approaches were used to fix the parameters:\\
  A1. Spectro-interferometric observations 
   were used to determine a wavelength-dispersed uniform disk diameter. A correction factor for hydrostatic
  C-stars was derived from hydrostatic models to convert the UD diameter to the Rosseland radius, which was then used 
  to determine the effective temperature $T_{\mathrm{eff}}$. \\
  A2. Spectroscopic measurements were compared
  to synthetic spectra from hydrostatic models to determine $T_{\mathrm{eff}}$ and C/O ratio. 
  The mass and log\,$g$ were constrained by comparing 
  spectro-interferometric observations with synthetic visibility profiles from hydrostatic models.\\
  The main advantage of approach 1 is the distance-independent determination of T$_\mathrm{eff}$. On the 
  other hand, conversion of the UD-radius to the Rosseland radius and the use of the apparent bolometric magnitude 
  introduces uncertainties. Approach 2 allows to constrain not only T$_\mathrm{eff}$, but also C/O, log\,$g$ and $M$. 
  One of the disadvantages of this technique is the unknown error that is introduced by the model. Additionally, the uncertainty in 
  distance, that is needed to constrain log\,$g$ and $M$, and the degeneracy between these two parameters, limits 
  the accuracy of the parameter determination. 
  This suggests that high-resolution spectroscopy is needed to fully discriminate between mass and log\,$g$.\\
  There is a very good agreement between the best-fitting hydrostatic model atmosphere and observations (interferometry, 
  spectroscopy and photometry). \\
  Our spectro-interferometric results are also an important tool to constrain and validate stellar AGB models, that are 
  still subject to severe uncertainties. We found that present TP-AGB tracks with a detailed treatment of molecular opacities 
  nicely reproduce the derived $T_{\rm eff}, L, {\rm C/O}$ values for TX Psc.


\begin{acknowledgements}
  The authors thank Angela Baier for fruitful discussions on ISO spectra and Walter Nowotny for helpful discussions on SEDs and 
  photometric filter systems. This work is supported by the Austrian Science Fund FWF under project number AP23006. 
 BA acknowledges support from Austrian Science Fund (FWF) Projects AP23006 \& AP23586 and from contract 
 ASI-INAF I/009/10/0. This research has made use of the SIMBAD database, operated at CDS, Strasbourg, France. We 
  acknowledge the variable star observations from the AAVSO International Database that were used in this research.
\end{acknowledgements}

\bibliographystyle{aa} 
\bibliography{klotz}

\end{document}